\documentclass[aps,prb,twocolumn,showpacs]{revtex4}

\usepackage{graphicx}
\usepackage{dcolumn}
\usepackage{bm}

\begin{document}


\title{Exchange and correlation effects in the relaxation of hot electrons in
noble metals}
\author{I.~G.~Gurtubay$^{1}$,
J.~M.~Pitarke$^{1,2}$, and P.~M.~Echenique$^{2,3}$}
\affiliation{
$^{1}$Materia Kondentsatuaren Fisika Saila, Zientzi Fakultatea,
Euskal Herriko Unibertsitatea, 644 Posta kutxatila, E-48080 Bilbo,
Basque Country,
Spain\\
$^{2}$Donostia International Physics Center (DIPC) and Centro Mixto
CSIC-UPV/EHU, Manuel de Lardizabal Pasealekua, E-20018 Donostia,
Basque Country, Spain\\
$^{3}$Materialen Fisika Saila, Kimika Fakultatea, Euskal Herriko
Unibertsitatea,1071 Posta kutxatila, E-20018 Donostia, Basque Country,
Spain}

\date{\today}

\begin{abstract}
We report extensive first-principles calculations of the inelastic lifetime of
low-energy electrons in the noble metals Cu, Ag, and Au. The quasiparticle
self-energy is computed with full inclusion of exchange and correlation (xc)
effects, in the framework of the $GW\Gamma$ approximation of many-body theory.
Although exchange and correlation may considerably reduce both the screening
and the bare interaction of hot electrons with the Fermi gas, these
corrections have opposite signs. Our results indicate that the overall effect
of short-range xc is small and $GW\Gamma$ linewidths are close to their
xc-free $G^0W^0$ counterparts, as occurs in the case of a free-electron gas.
\end{abstract}


\maketitle

\section{\label{sarrera}INTRODUCTION}

Relaxation lifetimes of excited electrons with energies larger than 1~eV
are mainly dominated by electron-electron (e-e) inelastic interactions of
the excited electron with the electrons in the valence bands of the
solid. The inelastic lifetime of these so-called hot electrons has been
investigated for many years on the basis of the free-electron gas (FEG)
or jellium description of the solid,\cite{QF,ritchie59,chemphys}
in which a homogeneous assembly of interacting electrons is
assumed to be immersed in a uniform positive background. Nonetheless,
time-resolved two-photon photoemission (TR-2PPE) experiments
\cite{schmu94,hertel96,ogawa97,cao97,aesch97,knoesel,cao98,petek}
and ballistic electron-emission spectroscopy (BEES)\cite{bees1}
have shown that band-structure effects play a key role in the decay
mechanism.

The first theoretical investigations of the hot-electron dynamics which
take into account explicitly the band structure of the solid were
performed only a few years ago.\cite{igorcu99,schoneal99} Since then,
first-principles calculations of the e-e scattering have been reported
for simple,\cite{almabecu00} noble, \cite{keyling00,igorau00,zhukovag01}
and transition metals.\cite{zhukov02,bacelar02,au-graz} Nevertheless, all
existing calculations have been performed within the $G^0W^0$
approximation of many-body theory,\cite{hedin} with no inclusion of
exchange and correlation (xc) effects.

In this work, we go beyond the $G^0W^0$ approximation by including xc
effects both in the description of the dynamical screening of the
many-electron system [we go beyond the random-phase approximation (RPA)
in the evaluation of the screened interaction $W$] and in the expansion
of the electron self-energy in terms of $W$. This is the $GW\Gamma$
approximation,\cite{mahan1,mahan2} which treats on the same footing xc
effects between pairs of electrons within the Fermi sea (screening
electrons) and between the hot excited electron and the Fermi sea. Mahan and
Sernelius\cite{mahan1} showed that the inclusion of the same vertex function
in the screened interaction and the numerator of the self-energy yields
results for the bandwidth of a homogeneous electron gas very similar to those
obtained in the $G^0W^0$ approximation, due to a large cancellation of vertex
corrections. Nonetheless, due to the well-known differences between the
dynamical response of the noble metals and the homogeneous electron gas, the
impact of exchange and correlation on the relaxation of hot electrons in these
materials required further investigation.

Our first-principles calculations start by solving self-consistently the
Kohn-Sham equations of density functional theory (DFT),\cite{kohn64,
kohn65} within the local density approximation (LDA) and with the use of
a plane-wave expansion of the single-particle Kohn-Sham Bloch states.
The electron-ion interaction is described by a nonlocal, norm-conserving
ionic pseudopotential,\cite{troumartins91} by keeping all $d$ electrons
as valence electrons. The single-particle Bloch states are then used to
compute the $GW\Gamma$ electron self-energy and hot-electron inelastic
lifetimes. For comparison, we also compute $G^0W^0$, $G^0W$, and
$GW^0\Gamma$ decay rates, with no inclusion of xc effects, with
inclusion of xc effects beyond the RPA in the screened interaction
$W$, and with inclusion of xc effects beyond the $G^0W^0$ in the expansion of
the
electron self-energy in terms of the RPA screened interaction $W^0$,
respectively. Our results indicate that the overall effect of short-range xc
is small and $GW\Gamma$ linewidths are close to their $G^0W^0$ counterparts,
as occurs in the case of a FEG.

In order to establish the role that occupied 
$d$ states play in the relaxation of hot electrons in the noble metals,
we also use a pseudopotential with all $d$ electrons assigned to the
core. We find that a major contribution from occupied $d$ states
participating in the screening of e-e interactions yields lifetimes of
low-energy excited hot electrons that are larger than
in the absence of $d$ states by a factor of $\sim 2$ in Cu and by a factor of  
$\sim 2.5$ in Ag and Au.

This paper is organized as follows. Explicit expressions for the electron decay
rate (inelastic lifetime broadening) in a FEG and periodic crystals are
derived in Sec.~\ref{theo}, in the $GW\Gamma$ approximation of many-body
theory. The results of numerical calculations of hot-electron lifetimes in the
noble metals Cu, Ag, and Au are presented in Sec.~\ref{results}. The summary
and conclusions are given in Sec.~\ref{sum}. Unless stated otherwise, atomic
units are used throughout, i. e., $e^2=\hbar=m_e=1$.

\section{\label{theo}THEORY}

Let us consider an arbitrary many-electron system of density $n_0({\bf
r})$. In the framework of many-body theory, the damping rate or
reciprocal lifetime of a quasiparticle in the single-particle state
$\phi_i({\bf r})$ of energy $\varepsilon_i$
($\varepsilon_i>\varepsilon_F$, $\varepsilon_F$ being the Fermi energy) is
obtained as the projection of the imaginary part of the
electron self-energy $\Sigma({\bf r},{\bf r}';\varepsilon_i)$ over
the quasiparticle-state itself:\cite{chemphys}
\begin{eqnarray}\label{eq:1}
\tau_i^{-1}=-{2}\int{\rm d}{{\bf r}}\int{\rm
d}{{\bf r}'}\,\phi_{i}^*({\bf r})\,{\rm Im}\,\Sigma({\bf r},{\bf
r}';\varepsilon_i)\,\phi_{i}({\bf r}').
\end{eqnarray}

Within many-body perturbation theory, it is possible to obtain the electron
self-energy as a series in the bare Coulomb interaction $v({\bf r},{\bf r}')$,
but due to the long range of this interaction such a perturbation series
contains divergent contributions. Therefore, the electron self-energy is
usually rewritten as a series in the frequency-dependent screened interaction
$W({\bf r},{\bf r}';\omega)$. To lowest order in the screened
interaction, the self-energy is obtained by integrating the product of
the interacting Green function $G({\bf r},{\bf r}',\varepsilon_i-\omega)$ and
the screened interaction $W({\bf r},{\bf r}';\omega)$, and is therefore called
the $GW$ self-energy. If one further replaces the interacting Green function
by its noninteracting counterpart $G^0({\bf r},{\bf
r}',\varepsilon_i-\omega)$, one finds the $G^0W$ self-energy and from
Eq.~(\ref{eq:1}) the following expression for the $G^0W$ lifetime broadening:
\begin{eqnarray}\label{eq:tau}
\tau_i^{-1}&=&-2\,\sum_f\int{\rm d}{{\bf r}}\int{\rm d}{{\bf
r}'}\,\phi_{i}^*({\bf r})\,\phi_{f}^*({\bf
r}')\cr\cr&\times&{\rm Im}\,W({\bf r},{\bf
r}';\varepsilon_i-\varepsilon_f)\,\phi_{i}({\bf r}')\,\phi_{f}({\bf
r}),
\end{eqnarray}
where the sum is extended over a complete set of
single-particle states $\phi_{f}({\bf r})$ of energy $\varepsilon_f$
($\varepsilon_F\le\varepsilon_f\le\varepsilon_i$). The screened
interaction $W({\bf r},{\bf r}';\omega)$ can be rigorously expressed as
follows
\begin{eqnarray}\label{eq:Wrpa}W({\bf r},{\bf
r}';\omega)&=&v({\bf r},{\bf r}')+\int{\rm d}{\bf r}_1\int{\rm d}{\bf
r}_2\,v({\bf r},{\bf r}_1)\cr\cr&\times&\chi({\bf r}_1,{\bf
r}_2;\omega)\,v({\bf r}_2,{\bf r}'),
\end{eqnarray}$\chi({\bf r},{\bf r};\omega)$ being the time-ordered
density-response function of the
many-electron system, which for the positive frequencies ($\omega>0$)
entering Eq.~(\ref{eq:tau}) coincides with the retarded density-response
function of linear-response theory. In the framework of
time-dependent DFT (TDDFT),\cite{tddft} the {\it exact} retarded
density-response function is obtained by solving the following integral
equation:\cite{Petersilka-96}
\begin{eqnarray}\label{eq:Xalda}&&\chi({\bf
r},{\bf r}';\omega)=\chi^0({\bf r},{\bf r}';\omega)+\int{\rm d}{\bf
r}_1\int{\rm d}{\bf r}_2\,\chi^0({\bf r},{\bf r}_1;\omega)\cr
\cr&&\times\left\{v({\bf r}_1,{\bf r}_2)+f^{xc}[n_0]({\bf r}_1,{\bf
r}_2;\omega)\right\}\chi({\bf r}_2, {\bf r}';\omega),\end{eqnarray}where
$\chi^0({\bf r},{\bf r}';\omega)$ denotes the density-response function
of noninteracting electrons
\begin{eqnarray}\label{eq:X0}\chi^0({\bf
r},{\bf r}';\omega)&=& 2\sum_{i,j}\frac
{f_i-f_j}{\varepsilon_i-\varepsilon_j+\omega+i\eta}\cr\cr
&\times&\phi_{i}({\bf r})\phi_{j}^*({\bf r})\phi_{j}({\bf
r}')\phi_{i}^*({\bf r}').
\end{eqnarray}
Here, $\phi_{i}({\bf r})$ and $\varepsilon_i$ denote the eigenfunctions and
eigenvalues of the Kohn-Sham hamiltonian of DFT, $f_i$ are Fermi-Dirac
occupation factors, $\eta$ is a positive infinitesimal, and the
frequency-dependent xc kernel $f^{xc}[n_0]({\bf r},{\bf r}';\omega)$ is the
functional derivative of the frequency-dependent xc potential $V_{xc}[n]({\bf
r},\omega)$ of TDDFT, to be evaluated at $n_0({\bf r})$:
\begin{equation}\label{fxc1}
f^{xc}[n_0]({\bf r},{\bf r}';\omega)=\left.{\delta V_{xc}[n]({\bf
r},\omega)\over\delta n({\bf r}',\omega)}\right|_{n=n_0}.
\end{equation}
In the RPA, $f_{xc}[n_0]({\bf r},{\bf r}';\omega)$ is set equal to zero and
Eq.~(\ref{eq:tau}) yields the so-called $G^0W^0$ lifetime
broadening.\footnote{This is the often-called $G^0W$-RPA.} In
the adiabatic LDA (ALDA),\footnote{Introduction of this xc kernel into
Eq.~(\ref{eq:Xalda}) yields the so-called $G^0W$-ALDA.}
\begin{equation}\label{fxc2}
f^{xc}({\bf r},{\bf r}';\omega)=\left.{dV_{xc}(n)\over
dn}\right|_{n=n_0({\bf r})}\,\delta({\bf r}-{\bf
r}'),
\end{equation}
$V_{xc}(n)$ being the static xc potential of a uniform
electron gas of density $n$.

The xc kernel $f^{xc}[n_0]({\bf r},{\bf r}';\omega)$, which is absent in the
RPA, accounts for the presence of an xc hole associated to all electrons in
the Fermi sea. Hence, one might be tempted to conclude that the full $G^0W$
approximation [with the formally exact screened interaction $W$ of
Eq.~(\ref{eq:Wrpa})] should be a better approximation than its $G^0W^0$
counterpart [with the screened interaction $W$ evaluated in the RPA]. However,
the xc hole associated to the excited hot electron is still absent in the
$G^0W$ approximation. Therefore, if one goes beyond RPA in the description of
$W$, one should also go beyond the $G^0W$ approximation in the expansion of the
electron self-energy in powers of $W$. By including xc effects both beyond RPA
in the description of $W$ and beyond $G^0W$ in the description of
the self-energy,\cite{mahan1,mahan2} the so-called $GW\Gamma$ approximation
yields a lifetime broadening that is of the $G^0W$ form [see
Eq.~(\ref{eq:tau})], but with the actual screened interaction $W({\bf r},{\bf
r}';\omega)$ replaced by a new effective screened interaction
\begin{eqnarray}\label{eq:Walda}
&&\tilde W({\bf r},{\bf r}';\omega)
=v({\bf r},{\bf r}')+\int{\rm d}{\bf r}_1\int{\rm
d}{\bf r}_2\,\left\{v({\bf r},{\bf
r}_1)\right.\cr\cr&&+\left.f^{xc}[n_0]({\bf r},{\bf
r}_1;\omega)\right\}\,\chi({\bf r}_1,{\bf r}_2;\omega)\,v({\bf r}_2,{\bf
r}'),
\end{eqnarray}
which includes all powers in $W$ beyond the $G^0W$ approximation.

\subsection{Free-electron gas}

In the case of a uniform FEG, there is translational invariance in all
directions, the single-particle states entering Eqs.~(\ref{eq:tau}) and
(\ref{eq:X0}) are momentum eigenfunctions $\phi_{\bf k}({\bf r})=\exp(i\,{\bf
k}\cdot{\bf r})$ of energy $\varepsilon_{\bf k}=k^2/2$, and Eq.~(\ref{eq:tau})
is easily found to yield
\begin{equation}\label{feg}
\tau_{\bf k}^{-1}=-2\int{d{\bf q}\over(2\pi)^3}\,{\rm Im}\,W({\bf
q},\varepsilon_{\bf k}-\varepsilon_{{\bf k}-{\bf q}}),
\end{equation}
the integral being subject to the condition
that $\varepsilon_F<\varepsilon_{{\bf k}-{\bf q}}<\varepsilon_{\bf
k}$, and $W({\bf q},\omega)$ being the Fourier transform of the
screened interaction $W({\bf r},{\bf r}';\omega)$ of Eq.~(\ref{eq:Wrpa}),
which in the $GW\Gamma$ approximation should be replaced by the
effective screened interaction $\tilde W({\bf r},{\bf r}';\omega)$
of Eq.~(\ref{eq:Walda}). We note that Eq.~(\ref{feg}) with $W({\bf q},\omega)$
replaced by the Fourier transform of $\tilde W({\bf r},{\bf r}';\omega)$
yields precisely the decay rate that one would obtain from the $GW\Gamma$
self-energy of Ref.~\onlinecite{mahan1}.

\subsection{Periodic crystals}

For periodic crystals, the single-particle states entering Eq.~(\ref{eq:tau})
are Bloch states $\phi_{{\bf k},i}({\bf r})$ and $\phi_{{\bf
k}-{\bf q},f}({\bf r})$ with energies $\varepsilon_{{\bf
k},i}$ and $\varepsilon_{{\bf k}-{\bf q},f}$, $i$ and $f$ denoting band
indices. Hence, Eq.~(\ref{eq:tau}) yields
\begin{eqnarray}\label{eq:tauG}
\tau_{{\bf k},i}^
{-1}&=&{1\over\pi^2}\sum_f\int_{\rm BZ}{{\rm d}{\bf q}}\sum_{{\bf
G},{\bf G}'}B_{{\bf k},i;{\bf k}-{\bf q},f}^*({\bf G})B_{{\bf k},i;{\bf
k}-{\bf q},f}({\bf G}')\cr\cr&\times&{\rm Im}\,W_{{\bf G},{\bf G}'}({\bf
q},\varepsilon_{{\bf k},i}-\varepsilon_{{\bf k}-{\bf
q},f}),
\end{eqnarray}
where the integral is extended over the first
Brillouin Zone (BZ), the vectors {\bf G} and {\bf G}' are
reciprocal-lattice vectors,
\begin{equation}\label{eq:Bif}
B_{{\bf k},i;{\bf k}-{\bf q},f}({\bf G})=\int{\rm d}{\bf r}\,\phi_{{\bf
k},i}^{\ast}({\bf r})\,{\rm e}^{{i}({\bf q+G})\cdot{\bf r}}\,\phi_{{\bf
k}-{\bf q},f}({\bf r}),
\end{equation}
and $W_{{\bf G},{\bf G}'}({\bf q},\omega)$ denote the Fourier coefficients of
the screened interaction $W({\bf r},{\bf r}';\omega)$, which are usually
expressed as follows
\begin{equation}\label{bands3}
W_{{\bf G},{\bf G}'}({\bf q},\omega)=
\epsilon_{{\bf G},{\bf G}'}^{-1}({\bf q},\omega)\,
v_{{\bf G}'}({\bf q})
\end{equation}
$v_{\bf G}({\bf q})=4\pi/|{\bf q}+{\bf
G}|^2$ being the Fourier transform of the bare Coulomb interaction, and
$\epsilon_{{\bf G},{\bf G}'}^{-1}({\bf q},\omega)$ being the so-called inverse
dielectric matrix.

We remark that for a given hot-electron energy $\varepsilon$ there are in
general various possible wave vectors and bands. Since $\tau^{-1}_{S{\bf
k},i}=\tau^{-1}_{{\bf k},i}$, $S$ denoting a point-group symmetry operation in
the crystal, one only needs to consider states inside the irreducible wedge 
of the BZ (IBZ). An energy-dependent reciprocal lifetime
$\tau^{-1}(\varepsilon)$ can  then be defined by doing a weighed  average over
all wave vectors and bands lying with the same energy in the IBZ.

\subsubsection{$G^0W^0$ approximation}

In the $G^0W^0$ approximation, the xc kernel
$f^{xc}[n_0]({\bf r},{\bf
r}';\omega)$ entering Eqs.~(\ref{eq:Xalda}) and (\ref{eq:Walda}) is set equal
to zero. Hence, in
this approximation the dielectric matrix is
\begin{equation}\label{bands4}\epsilon_{{\bf G},{\bf G}'}({\bf
q},\omega)=\delta_{{\bf G},{\bf G}'}-v_{{\bf G}}({{\bf q}})\,\chi_{{\bf G},{\bf
G}'}^0({\bf q},\omega),
\end{equation}
$\chi_{{\bf G},{\bf G}'}^0({\bf q},\omega)$ being the Fourier coefficients of
the noninteracting density-response function of Eq.~(\ref{eq:X0}):
\begin{eqnarray}\label{eq:X0GG}
\chi_{{\bf G},{\bf
G}'}^0({\bf q},\omega)&=&2\,\int_{BZ}\frac{{\rm d}{\bf
k}}{(2\pi)^3}\,\sum_{n,n'}\,{f_{{\bf k},n}-f_{{\bf k}+{\bf
q},n'}\over\epsilon_{{\bf k},n}-\epsilon_{{\bf k}+{\bf
q},n'}+\omega+{i}\eta}\cr\cr&\times&\langle\phi_{{\bf k},n}|e^{-{
i}({\bf q}+{\bf G})\cdot{\bf r}}|\phi_{{\bf k}+{\bf
q},n'}\rangle\cr\cr&\times&\langle\phi_{{\bf k}+{\bf q},n'}|e^{{i}({\bf
q}+{\bf G}')\cdot{\bf r}}|\phi_{{\bf
k},n}\rangle.
\end{eqnarray}

Couplings of the wave vector ${\bf q}+{\bf G}$ to wave vectors ${\bf q}+{\bf
G}'$ with ${\bf G}\neq{\bf G}'$ appear
as a consequence of the existence of electron-density variations in the
solid. If these terms, representing the so-called crystalline local-field
effects (LFE), are neglected, one can write Eq.~(\ref{eq:tauG}) as
follows
\begin{eqnarray}\label{eq:taudiag}
\tau_{{\bf
k},i}^{-1}&=&{1\over\pi^2}\sum_f\int_{\rm BZ}{{\rm d}{\bf q}}\sum_{\bf
G}{\left|B_{{\bf k},i;{\bf k}-{\bf q},f}({\bf
G})\right|^2\over\left|{\bf q}+{\bf G}\right|^2}\cr\cr&\times&{{\rm
Im}\left[\epsilon_{{\bf G},{\bf G}}({\bf q},\varepsilon_{{\bf
k},i}-\varepsilon_{{\bf k}-{\bf q},f})\right]\over|\epsilon_{{\bf G},{\bf
G}}({\bf q},\varepsilon_{{\bf k},i}-\varepsilon_{{\bf k}-{\bf
q},f})|^2}.
\end{eqnarray}
This expression accounts explicitly for the three main ingredients entering the
hot-electron decay process. First of all, the coupling of the hot electron
with available states above the Fermi level is dictated by the matrix elements
$B_{{\bf k},i;{\bf k}-{\bf q},f}({\bf G})$. Secondly, the imaginary part of
the dielectric matrix $\epsilon_{{\bf G},{\bf G}}({\bf q},\varepsilon_{{\bf
k},i}-\varepsilon_{{\bf k}-{\bf q},f})$ represents a measure of the
number of states available for the creation of e-h pairs with momentum
and energy ${\bf q}+{\bf G}$ and $\varepsilon_{{\bf k},i}-\varepsilon_{{\bf
k}-{\bf q},f}$, respectively. Thirdly, the dielectric matrix in the
denominator accounts for the many-body e-e interactions in the Fermi sea,
which dynamically screen the interaction with the external hot electron.

\subsubsection{$G^0W$ approximation}

In the $G^0W$ approximation, the Fourier coefficients $W_{{\bf G},{\bf
G}'}({\bf q},\omega)$ still take the form of Eq.~(\ref{bands3}), but with the
inverse dielectric matrix now given by the following general expression:
\begin{equation}\label{eq:WGG1}
\epsilon_{{\bf G},{\bf G}'}^{-1}({\bf q},\omega)=\delta_{{\bf G},{\bf
G}'}+v_{{\bf G}}({{\bf q}})\,\chi_{{\bf G},{\bf G}'}({\bf q},\omega),
\end{equation}
where $\chi_{{\bf G},{\bf G}'}({\bf q},\omega)$ are the Fourier coefficients of
the interacting density-response function of Eq.~(\ref{eq:Xalda}):
\begin{eqnarray}\label{eq:XGG}
&&\chi_{{\bf G},{\bf
G}'}({\bf q},\omega)=\chi^0_{{\bf G},{\bf G}'}({\bf q},\omega)+
\sum_{{\bf G}_1,{\bf G}_2}\,\chi^0_{{\bf G},{\bf G}_1}({\bf
q},\omega)\cr\cr&&\times\left\{v_{{\bf G}_1}({{\bf q}})+f_{{\bf
G}_1,{\bf G}_2}^{xc}[n_0]({\bf q},\omega)\right\}\chi_{{\bf G}_2,{\bf
G}'}({\bf q},\omega),
\end{eqnarray}
$f_{{\bf G},{\bf G}'}^{xc}[n_0]({\bf
q},\omega)$ being the Fourier coefficients of the xc kernel
of Eq.~(\ref{fxc1}). In the ALDA,
\begin{equation}\label{fxcgg}
f_{{\bf G},{\bf G}'}^{xc}({\bf
q},\omega)=\int d{\bf r}\,{\rm e}^{-i({\bf G}-{\bf G}')\cdot{\bf
r}}\,\left.{dV_{xc}(n)\over
dn}\right|_{n=n_0({\bf r})}.
\end{equation}

\subsubsection{$GW\Gamma$ approximation}

In the $GW\Gamma$ approximation, the lifetime broadening is
still of the form of Eqs.~(\ref{eq:tauG})-(\ref{bands3}), but with the
test-charge--test-charge inverse dielectric matrix of Eq.~(\ref{eq:WGG1})
replaced by the new test-charge--electron inverse dielectric matrix
\begin{eqnarray}\label{eq:WGG2}
&&\tilde\epsilon_{{\bf G},{\bf G}'}^{-1}({\bf q},\omega)=
\delta_{{\bf G},{\bf G}'} +\sum_{{\bf G}''} 
\left\{ v_{{\bf G}}({{\bf q}})\,\delta_{{\bf G},{\bf G}''}\right.\cr\cr
&&+\left.f_{{\bf G},{\bf G}''}^{xc}[n_0]({\bf q},\omega)\right\}
\,\chi_{{\bf G}'',{\bf G}'}({\bf q},\omega).
\end{eqnarray}
where $\chi_{{\bf G},{\bf G}'}({\bf q},\omega)$ are the Fourier coefficients of
Eq.~(\ref{eq:XGG}). If one sets the xc kernel
$f_{{\bf G},{\bf G}'}^{xc}[n_0]({\bf q},\omega)$ entering Eqs.~(\ref{eq:XGG})
and (\ref{eq:WGG2}) equal to zero, one finds the $G^0W^0$ approximation.
Instead, if one only sets the xc kernel $f_{{\bf G},{\bf G}'}^{xc}[n_0]({\bf
q},\omega)$ entering Eq.~(\ref{eq:XGG}) equal to zero, one finds the
$GW^0\Gamma$ approximation.

\section{\label{results}Results and Discussion}

In the calculations presented in this section, all the single-particle Bloch
states and energies entering Eqs.~(\ref{eq:tauG}), (\ref{eq:Bif}), and
(\ref{eq:X0GG}) are taken to be the eigenfunctions and eigenvalues of the
LDA Kohn-Sham hamiltonian of DFT. We expand the single-particle Bloch states in
a plane-wave basis, we invoke the LDA with the Perdew-Zunger
parametrization\cite{perdewzunger} of the Quantum Monte Carlo uniform-gas xc
energy of
Ceperly and Alder\cite{ceperlyalder}, and we describe the electron-ion
interaction with the use of a nonlocal, norm-conserving ionic
pseudopotential.\cite{troumartins91}

Cu, Ag, and Au are noble metals with entirely filled $nd$ bands, $n$ being 3,
4, and 5, respectively. Slightly below the Fermi level [at
$\varepsilon-\varepsilon_F\sim 2\,{\rm eV}$ in Cu and Au, and at
$\varepsilon-\varepsilon_F\sim 4\,{\rm eV}$ in Ag] there are $d$ bands capable
of holding 10 electrons per atom. The one remaining $(n+1)s$ electron per atom
occupies a free-electron-like band below and above the $d$ bands. Hence, a
combined description of the localized $nd^{10}$ and delocalized $(n+1)s^1$
electrons is needed in order to address the actual electronic response
of these materials.

The results presented below have been found by either
keeping all $nd^{10}$ and $(n+1)s^1$ electrons (full calculation) or
keeping only the $(n+1)s^1$ electrons ($s$-calculation) as valence electrons
in the generation of the pseudopotential. The full calculation has
required a kinetic-energy cut-off as large as $75\,{\rm
Ry}$.\footnote{For Cu, we keep $\sim 900$ plane waves in the expansion of
each Bloch state. In the case of Ag and Au, we keep $\sim 1300$ plane
waves.} Well-converged results have been found for all hot-electron
energies under study ($0.5\,{\rm eV}\leq\varepsilon-\varepsilon_F\leq 3.5\,{\rm
eV}$), with the inclusion of conduction bands up to a maximum energy of
$\sim 25\,{\rm eV}$ above the Fermi level. Samplings of the BZ have been
performed on a 20$\times$20$\times$20 mesh containing 256 points in the
IBZ, although well-converged results have sometimes been obtained on
16$\times$16$\times$16 meshes. Reciprocal-space sums have been extended
over 15 {\bf G} vectors, the magnitude of the maximum momentum transfer
{\bf q}+{\bf G} being well over the upper limit of $\sim 2q_F$, $q_F$
being the Fermi momentum.

\subsection {Copper}

\begin{figure}\vspace*{1cm}
\includegraphics[width=0.95\linewidth]{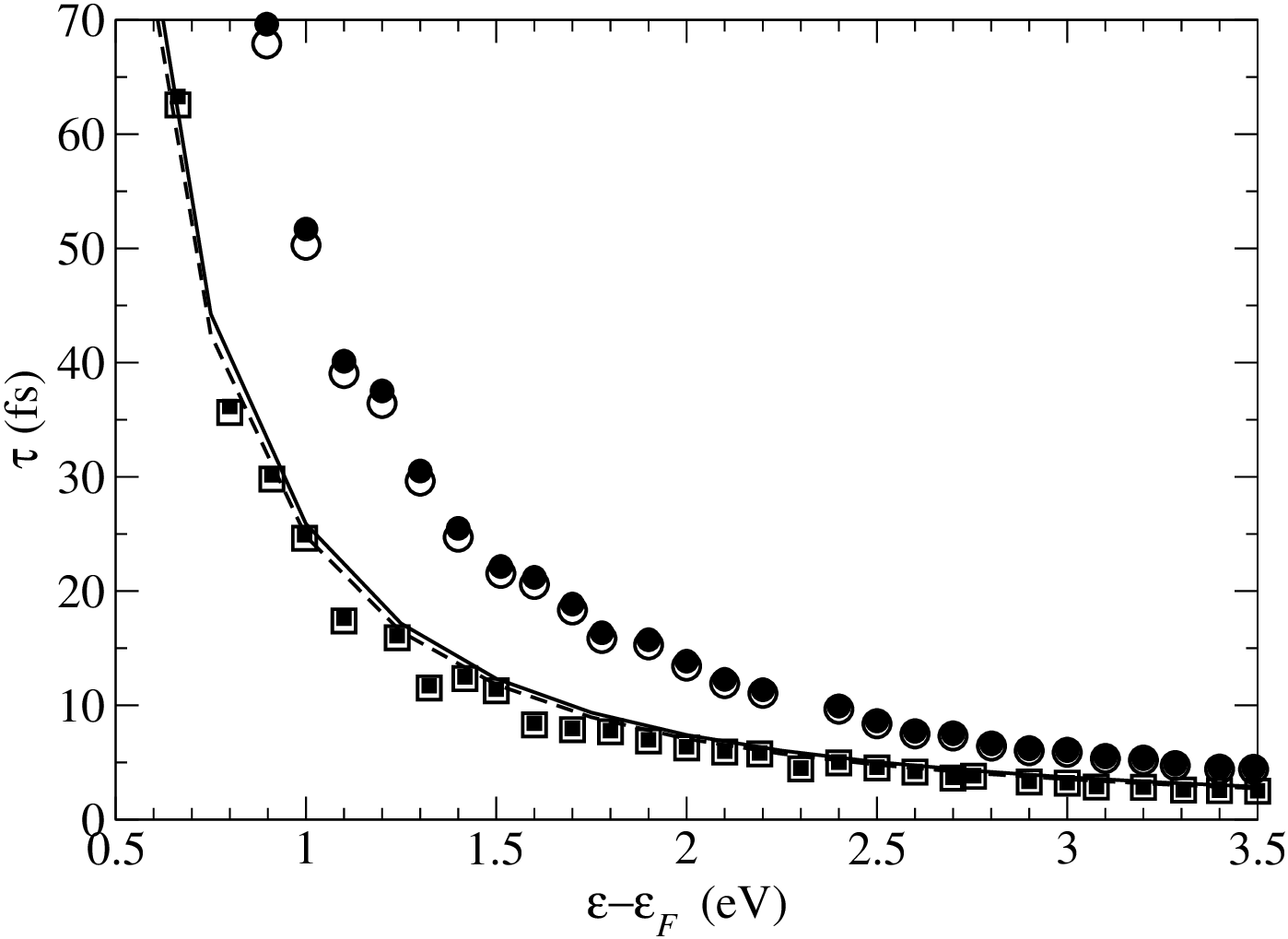}
\caption{\label{fig:cudslfe}
Average lifetime of hot electrons in Cu, as a function of the hot-electron
energy $\varepsilon-\varepsilon_F$ with respect to the Fermi level. Solid and
open circles represent full $GW\Gamma$ and $G^0W^0$ calculations, as obtained
from Eqs.~(\ref{eq:tauG})-(\ref{bands3}) with the test-charge--electron
inverse dielectric matrix of Eq.~(\ref{eq:WGG2}) and with the RPA dielectric
matrix of Eq.~(\ref{bands4}), respectively. Solid and open squares represent
the corresponding $GW\Gamma$ and $G^0W^0$ $s$-calculations, respectively,
where the $3d$-shell is assigned to the core in the generation of the
pseudopotential. Solid and dashed lines represent $GW\Gamma$ and $G^0W^0$ FEG
calculations, as obtained from Eq.~(\ref{feg}) with the electron-density
parameter $r_s=2.67$ [$r_s=(3/4\pi n_0)^{1/3}$] corresponding to the average
density $n_0$ of $4s^1$ electrons in Cu. The Fourier coefficients
$f_{{\bf G},{\bf G}'}^{xc}[n_0]({\bf q},\omega)$ and
$f^{xc}[n_0]({\bf q},\omega)$ entering the crystal and FEG calculations have
both been calculated in the ALDA with use of the Perdew-Zunger xc potential of
a uniform electron gas.}
\end{figure}

Fig.~\ref{fig:cudslfe} shows our full first-principles $GW\Gamma$ calculation
of the average lifetime $\tau(E)$ of hot electrons in Cu (solid circles), as
obtained in the ALDA from Eqs.~(\ref{eq:tauG})-(\ref{bands3}) but with the
inverse dielectric matrix $\epsilon_{{\bf G},{\bf G}'}^{-1}({\bf q},\omega)$
replaced by that of Eq.~(\ref{eq:WGG2}). For comparison, we also plot in the
same figure our full first-principles $G^0W^0$ calculation (open circles),
which reproduces
previous calculations,\cite{igorcu99} first-principles $GW\Gamma$ and $G^0W^0$
$s$-calculations (solid and open squares), and $GW\Gamma$ and $G^0W^0$ FEG
calculations with the electron density $n_0$ equal to that of $4s^1$ electrons
in Cu (solid and dashed lines).

We note from Fig.~\ref{fig:cudslfe} that xc effects yield hot-electron
lifetimes that are larger than in the absence of exchange and correlation by
no more than $\sim 3\%$ in the whole hot-electron energy range under study, as
occurs
when $d$ electrons are assigned to the core ($s$-calculations) and in the case
of a FEG. Exchange-correlation effects included in the $GW\Gamma$ scheme have
two sources, as discussed in Sec.~\ref{theo}. Firstly, there is the
reduction of the screening due to the presence of an xc hole associated
to occupied states below the Fermi level, as in the $G^0W$ approximation.
Secondly, there is the xc hole associated to the excited hot electron over the
Fermi level, as in the $GW^0\Gamma$ approximation. These contributions have
opposite signs and it is the latter which dominates.

\begin{figure}
\vspace* {1cm}
\includegraphics[width=0.95\linewidth]{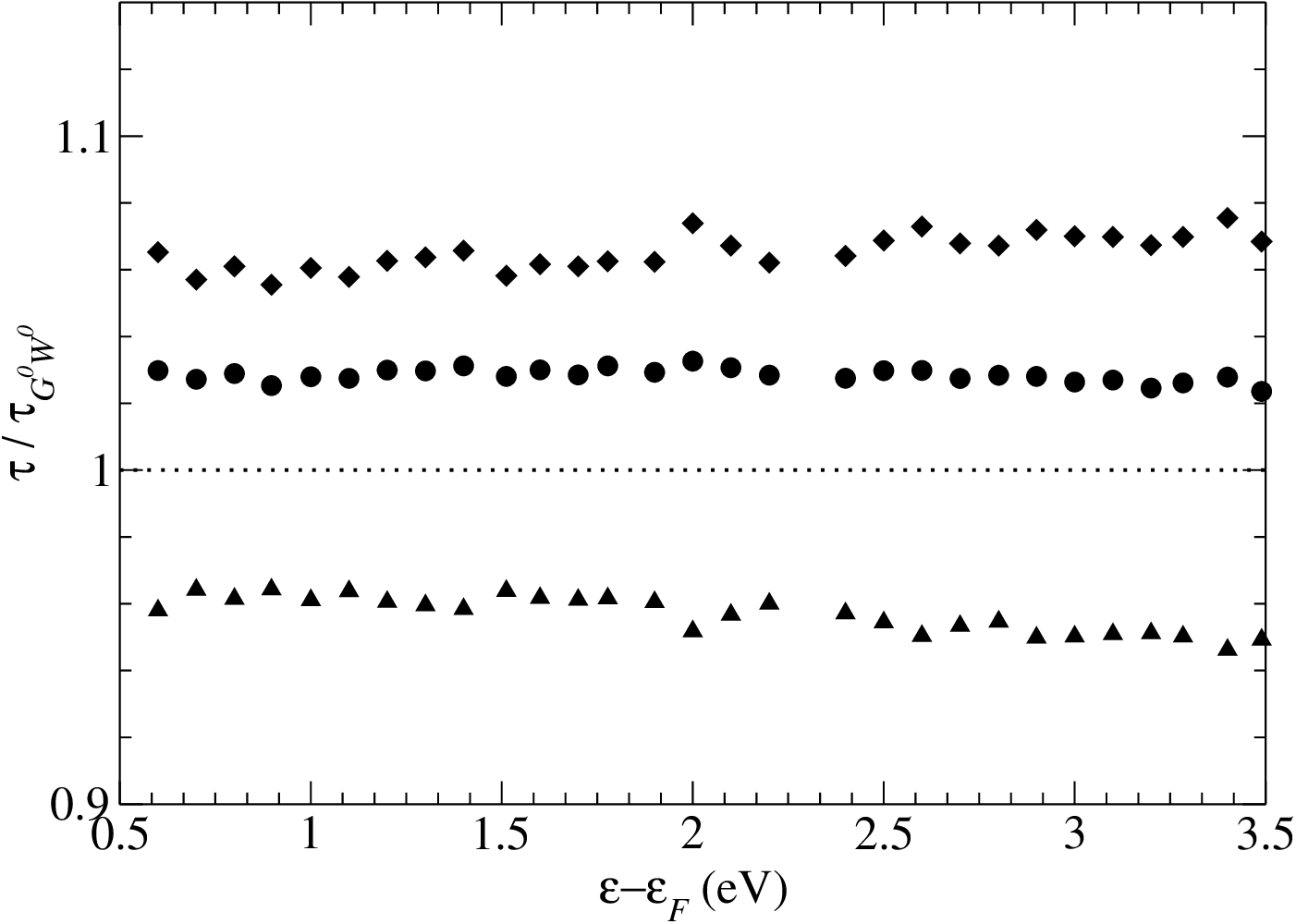}
\\
\vspace* {1.2cm}
\includegraphics[width=0.95\linewidth]{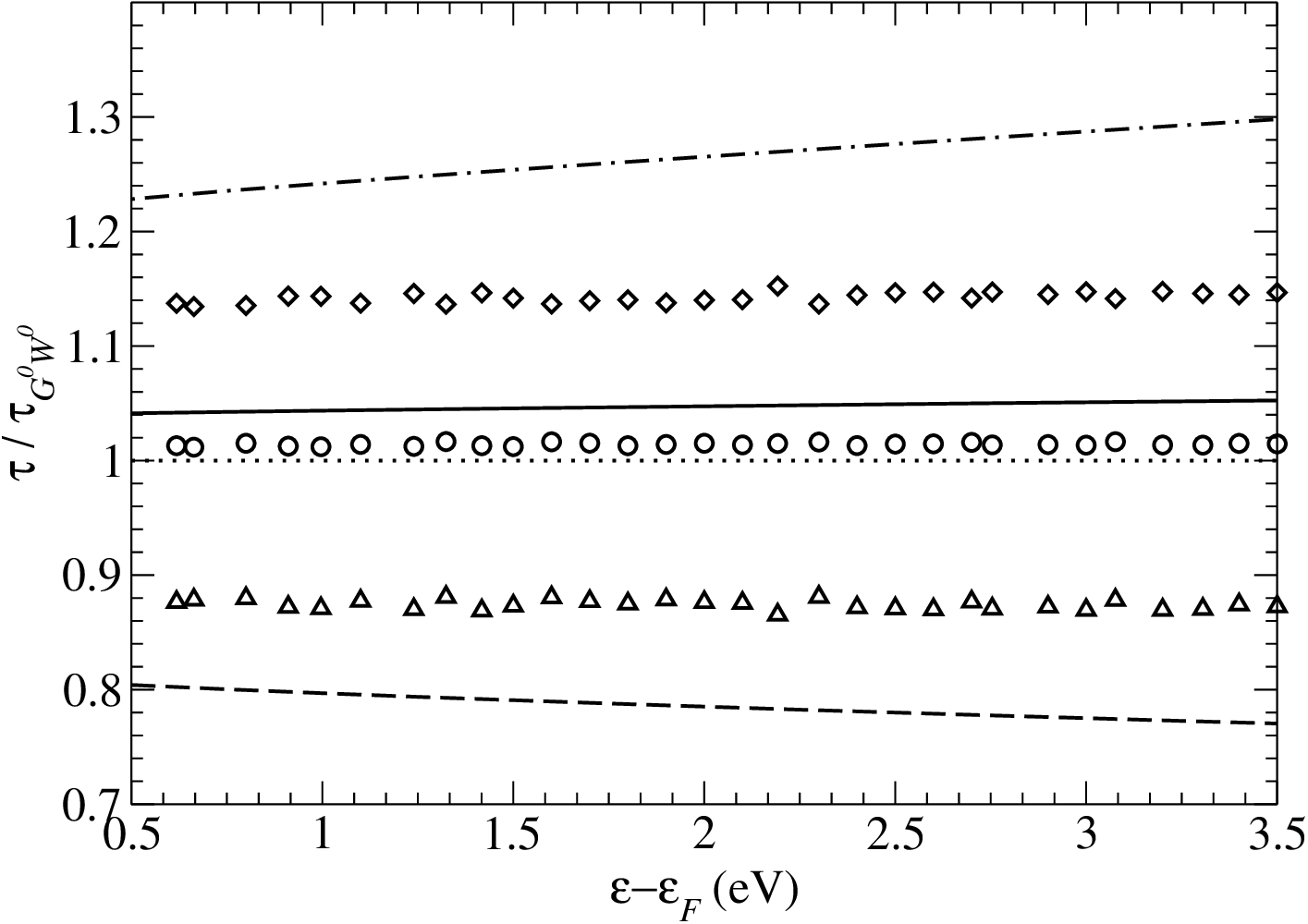}
\caption{\label{fig:cualda}
Impact of exchange and correlation on the lifetime of hot electrons in Cu, as a
function of the hot-electron energy $\varepsilon-\varepsilon_F$ wit respect to
the Fermi level. Triangles, diamonds and circles represent the ratios
$\tau_{G^0W}/\tau_{G^0W^0}$, $\tau_{GW^0\Gamma}/\tau_{G^0W^0}$, and
$\tau_{GW\Gamma}/\tau_{G^0W^0}$ in the presence (upper panel) and in the
absence (lower panel) of $3d^{10}$ occupied states, respectively. Dashed,
dashed-dotted and solid lines represent the corresponding ratios for a FEG
with $r_s=2.67$.}
\end{figure}

The impact of xc effects on the lifetime of hot electrons in Cu is illustrated
in Fig.~\ref{fig:cualda}, where $G^0W$, $GW^0\Gamma$, and $GW\Gamma$
calculations are compared to their $G^0W^0$ counterparts, with (upper panel)
and without (lower panel) inclusion of $d$ states. The impact of exchange and
correlation on either screening or the bare interaction of the hot electron
with the Fermi sea, which is significant in the absence of $d$ states (lower
panel), is considerably reduced in the presence of $d$ electrons. Moreover,
these small xc effects partially compensate each other, leading to an overall
effect of no more than $3\%$.

Now we focus on the role that localized $d$ bands play in the decay mechanism
of hot electrons in Cu. This issue has been investigated before,\cite{igorcu99}
by replacing the various contributions to Eq.~(\ref{eq:taudiag})
by the corresponding FEG contributions. Here, we follow a more meaningful
procedure. As in Ref.~\onlinecite{igorcu99} we use
Eq.~(\ref{eq:taudiag}) (hence, with inclusion of neither XC effects nor 
crystalline local-field corrections), but now we replace the various
contributions
to the full calculation
by the corresponding $s$-contributions where the $3d$ shell is assigned to the
core. These
contributions are: (i) the imaginary part of the dielectric matrix
$\epsilon_{{\bf G},{\bf G}}({\bf q},\omega)$, which represents
a measure of both the number of states available for the creation of e-h pairs
and the coupling between states below and above the Fermi level, (ii)
the denominator $\left|\epsilon_{{\bf G},{\bf G}}({\bf q},\omega)\right|^2$,
which accounts for screening effects, and (iii) the matrix elements
$B_{{\bf k},i;{\bf k}-{\bf q},f}({\bf G})$ accounting for the coupling of the
hot electron with available states above the Fermi level.

\begin{figure}
\vspace* {1cm}
\includegraphics[width=0.95\linewidth]{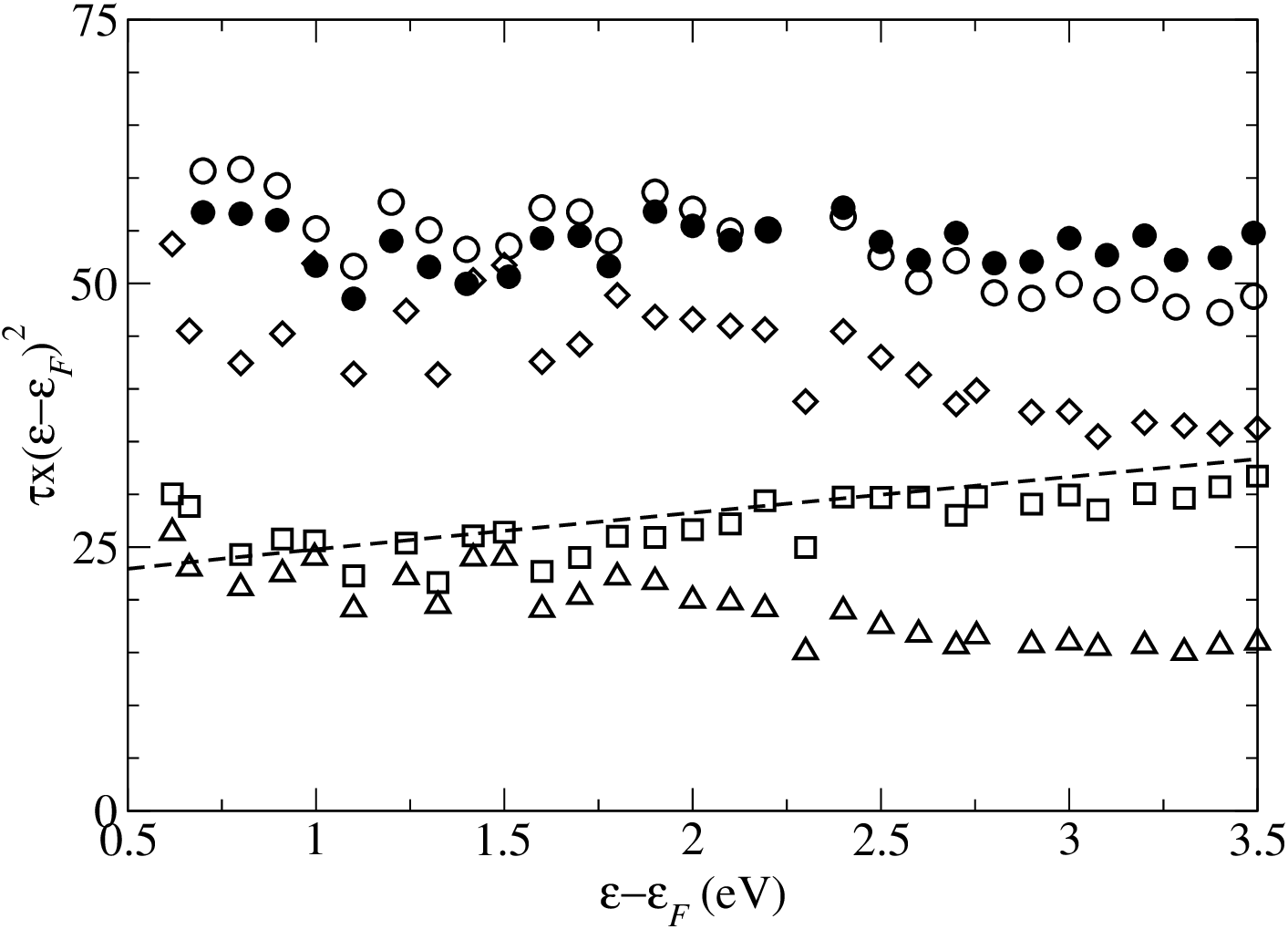}
\caption{\label{fig:cumix}
Scaled lifetimes $\tau(\varepsilon)\times (\varepsilon-\varepsilon_F)^2$ of hot
electrons in Cu. Open circles, squares, triangles, and diamonds represent
$G^0W^0$ calculations obtained from Eq.~(\ref{eq:taudiag}) (hence, with no
inclusion of crystalline local-field effects), (i) in the presence of $3d$
states (open circles), (ii) in the absence of $3d$ states (squares), (iii)
including $3d$ states only in the evaluation of the imaginary part of
$\epsilon_{{\bf G},{\bf G}}({\bf q},\omega)$ (triangles), and (iv) including
$3d$ states in the evaluation of the imaginary part of
$\epsilon_{{\bf G},{\bf G}}({\bf q},\omega)$
as well as  in the evaluation of the denominator $\left|\epsilon_{{\bf
G},{\bf G}}({\bf q},\omega)\right|^2$ (diamonds). Solid circles represent the
full $GW\Gamma$ matricial calculation of Fig.~\ref{fig:cudslfe}, but now
multiplied by
$(\varepsilon-\varepsilon_F)^2$. The dashed line represents the $G^0W^0$
calculation for hot electrons in a FEG with $r_s=2.67$.} 
\end{figure}

Fig.~\ref{fig:cumix} shows first-principles $G^0W^0$ calculations of the scaled
lifetime $\tau(\varepsilon)\times (\varepsilon-\varepsilon_F)^2$ of hot
electrons in Cu, as obtained from Eq.~(\ref{eq:taudiag}). We start with the
$s$-calculation (open squares), which we obtain by assigning the
$3d$ shell to the core in the generation of the pseudopotential. This
calculation nearly coincides with the $G^0W^0$ lifetime broadening of hot
electrons in a free-gas of $4s^1$ electrons (dashed line), which must be a
consequence of the fact that band-structure effects are almost entirely due to
the presence of occupied $d$ bands. In Cu, localized $d$ bands contribute to
the decay of $sp$ hot electrons by either opening a $d$-band scattering
channel at $\sim 2\,{\rm eV}$ or by screening the e-e interactions. While the
opening of the $d$-band scattering channel modifies the imaginary part of the
dielectric matrix $\epsilon_{{\bf G},{\bf G}}({\bf q},\omega)$, the screening
of $d$ electrons modifies the denominator
$\left|\epsilon_{{\bf G},{\bf G}}({\bf q},\omega)\right|^2$. Hence, we have
gone beyond the $s$-calculation (open squares) by first including all $d$
states only in the evaluation of ${\rm Im}\left[\epsilon_{{\bf G},{\bf
G}}({\bf q},\omega)\right]$ (triangles) and then including all $d$ states
also in the evaluation of $\left|\epsilon_{{\bf G},{\bf G}}({\bf
q},\omega)\right|^{-2}$ (diamonds). These calculations clearly
indicate that: (i) occupied $d$ bands yield an slightly enhanced hot-electron
decay (reduced lifetime) at the opening of the $d$-band scattering channel
($\varepsilon-\varepsilon_F>2\,{\rm eV}$), which is considerably smaller than
expected from the greatly enhanced density of states at these energies due to
the small coupling between $d$ states below and $sp$ states above the Fermi
level for
the creation of e-h pairs, and (ii) the key role that $d$ electrons play in
the hot-electron decay is mainly due to screening effects. We also note that
the full diagonal $G^0W^0$ calculation (open circles), which includes neither
crystalline local-field corrections nor xc effects, is very close to the full
$GW\Gamma$ calculation (solid circles). This shows that the overall impact of
these effects in this material is never larger than a few per cent. 

The screening of $d$ electrons in the noble metals was included by
Quinn,\cite{quinn-63} by embedding the $ns^1$ free electrons in a polarizable
background of $d$ electrons characterized by a dielectric constant
$\epsilon_d$ instead of unity. The corrected lifetime is then found to be
larger than in the absence of $d$ electrons by roughly a factor of
$\epsilon_d^{1/2}$. For Cu and Au $\epsilon_d=5.6$, and for Ag
$\epsilon_d=3.4$.\cite{jc} Hence, this simple model yields corrected lifetimes
in Cu
that are larger than in a FEG by roughly a factor of $\sim 2.5$, in
qualitative agreement with first-principles calculations. Nevertheless, this
model cannot account for the existing differences between the impact of
occupied $d$ states in the lifetime broadening of hot electrons in Cu and Au,
which both have approximately the same value of $\epsilon_d$.

A comparison between $G^0W^0$ hot-electron lifetimes in Cu and those determined
from TR-TPPE experiments was presented in Ref.~\onlinecite{igorcu99}. At
low hot-electron energies, the calculated lifetimes were found to be in
agreement with the low-energy measurements of Knoesel {\it et
al.}.\cite{knoesel} At larger energies, very good agreement was found with the
lifetimes measured by Ogawa {\it et al.}\cite{ogawa97} at the (110) surface of
Cu, the only low-index surface with no band gap for electrons emitted in the
direction perperdincular to the surface. Our results indicate that the
inclusion of exchange and correlation does not substantially change this
agreement.

\subsection{Silver}

\begin{figure}
\vspace* {1cm}
\includegraphics[width=0.95\linewidth]{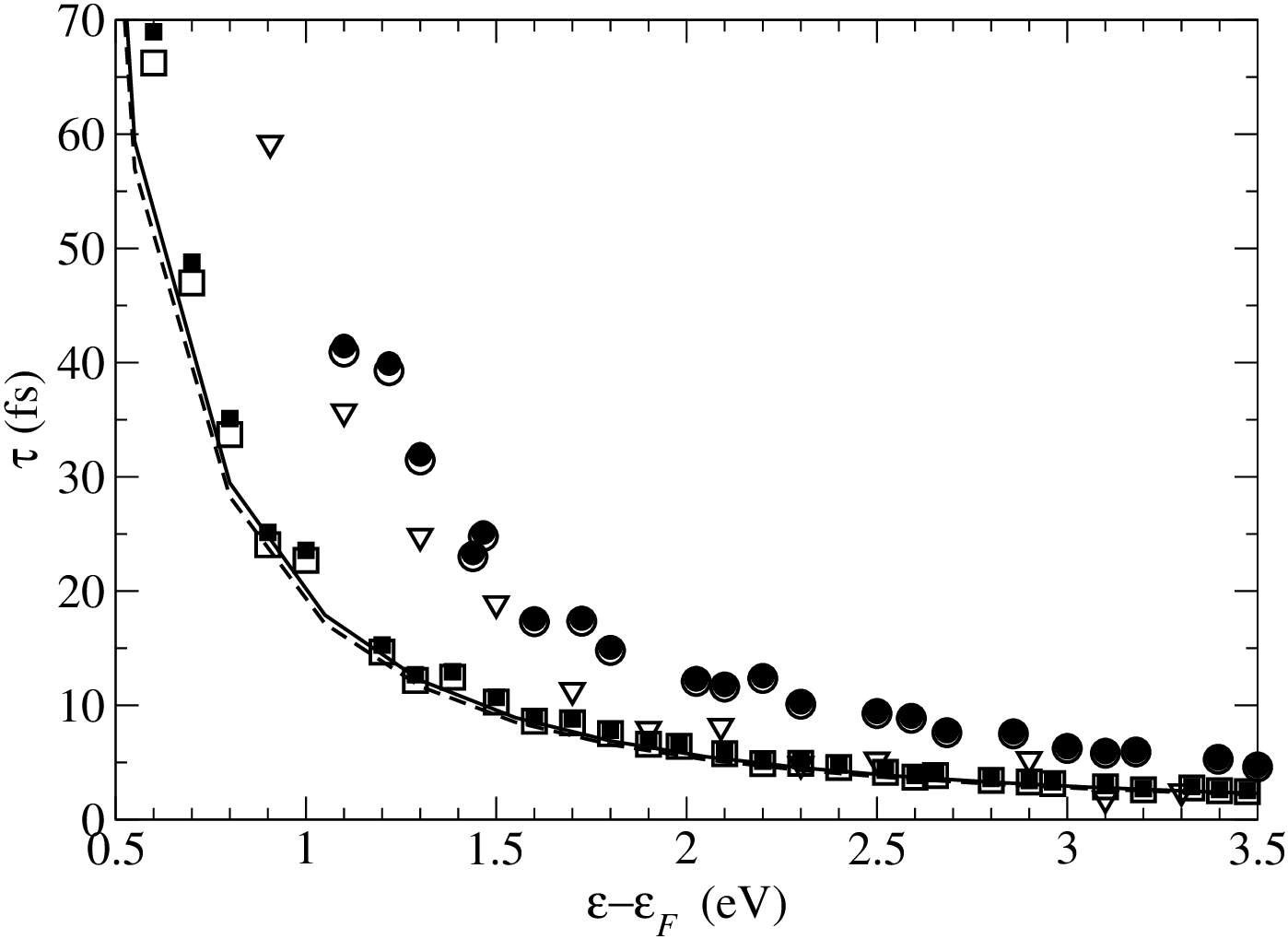}
\caption{\label{silver1}
As in Fig.~\ref{fig:cudslfe}, but for Ag. Here, solid and dashed lines
represent  
$GW\Gamma$ and $G^0W^0$ lifetimes of hot electrons in a FEG with $r_s=3.02$,
corresponding to the average density of $5s^1$ electrons in Ag. Inverted
triangles represent the TR-TPPE measurements reported in
Ref.~\onlinecite{aes02}.}
\end{figure}

We have plotted in Fig.~\ref{silver1} our full first-principles $GW\Gamma$
calculation of the average lifetime $\tau(\varepsilon)$ of hot electrons 
in Ag (solid circles), as
obtained in the ALDA from Eqs.~(\ref{eq:tauG})-(\ref{bands3})
 but with the inverse dielectric
matrix $\epsilon_{{\bf G},{\bf G}'}^{-1}({\bf q},\omega)$ replaced by that of
Eq.~(\ref{eq:WGG2}). For comparison, we also plot in the same figure our full
first-principles $G^0W^0$ calculation (open circles), first-principles
$GW\Gamma$ and $G^0W^0$ $s$-calculations (solid and open squares), and
$GW\Gamma$ and $G^0W^0$ FEG calculations with the electron density $n_0$ equal
to that of $5s^1$ electrons in Ag (solid and dashed lines). 
Also shown in this figure are the TR-2PPE measurements reported
by Bauer and Aeschlimann \cite{aes02} for the lifetime of hot 
electrons in a thin Ag film (inverted triangles).

Fig.~\ref{silver1} shows that both in the presence (solid and open
circles) and in the absence (solid and open squares) of $d$ states the
overall impact of exchange and correlation on the lifetime of hot
electrons in Ag is very small ($\sim 2\%$) in the whole hot-electron
energy range under study. As occurs in the case of Cu, the two separate
sources of xc effects, which are both found to yield significant effects
in the absence of $d$ states and considerably smaller effects in the
presence of $d$ states, almost cancel each other. We also
note from Fig.~\ref{silver1} that at the lowest hot-electron energies
there is good agreement between our full $G^0W^0$ and $GW\Gamma$
calculations and the experimental data, which are both $\sim 2.5$ larger
than in the absence of occupied $d$ states. At larger energies, however, our
calculations lie slightly higher than the experimental curve.

At this point, it is interesting to notice that while hot-electron
lifetimes in a FEG with $r_s=2.67$ (corresponding to the average
density of $4s^1$ electrons in Cu) are larger than in a FEG with
$r_s=3.01$ (corresponding to the average density of $5s^1$ electrons in
Ag), hot electrons are found to live slightly longer in Ag than in Cu. 
Both in Cu and Ag first-pinciples
$s$-calculations nearly coincide with their FEG counterparts, showing
that band-structure effects are almost entirely due to the presence of
$d$ electrons. Hence, the impact of $d$ electrons is found to be larger
in Ag (with the onset of the $d$ band at $\sim 4\,{\rm eV}$ below the
Fermi level $\varepsilon_F$) than in Cu (with the onset of the $d$ band
at $\sim 2\,{\rm eV}$ below $\varepsilon_F$).

\begin{figure}
\vspace* {1cm}
\includegraphics[width=0.95\linewidth]{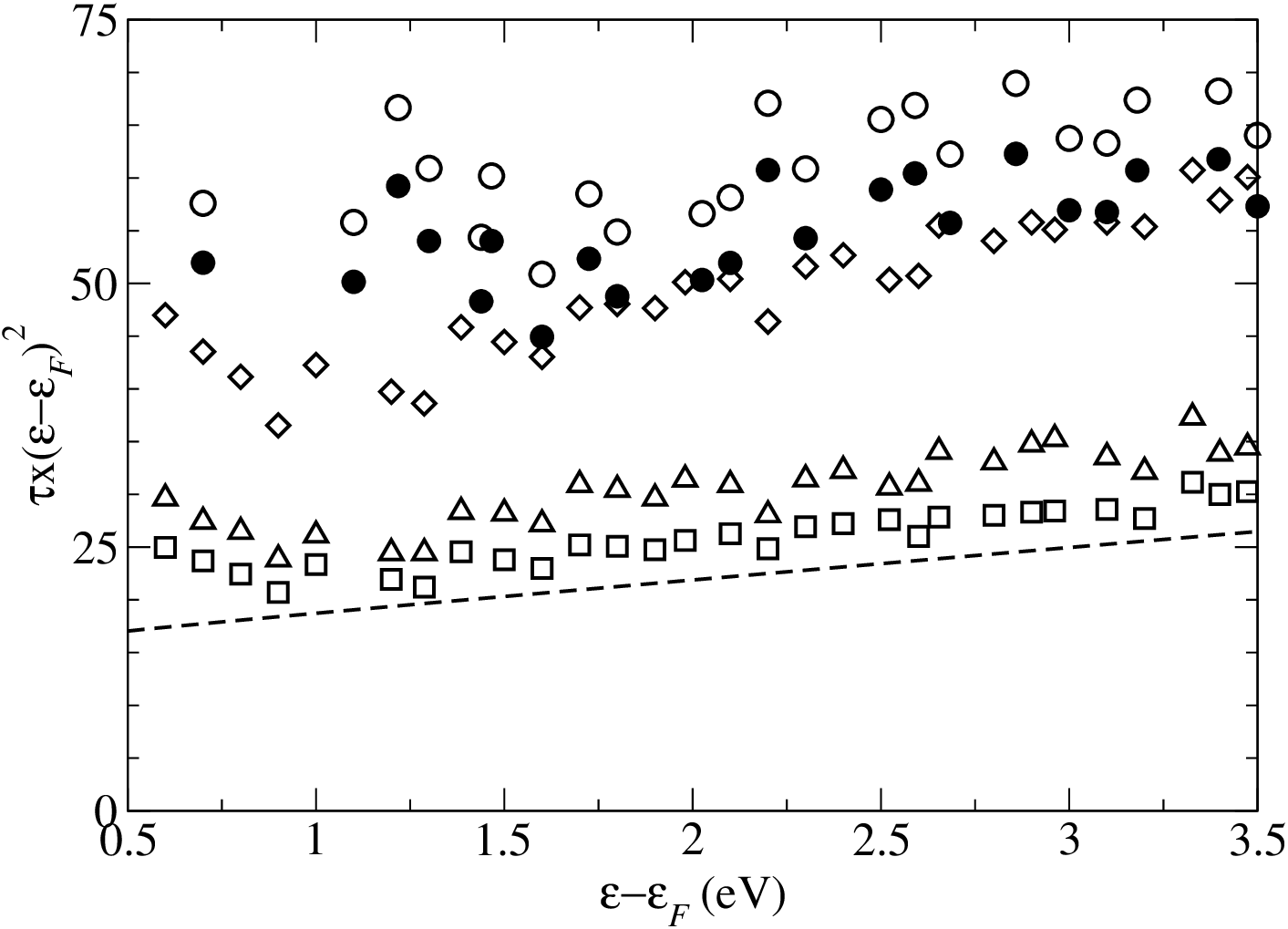}
\caption{\label{silver2}
As in Fig.~\ref{fig:cumix}, but for Ag. Solid circles now represent the
full $GW\Gamma$ matricial calculation of Fig.~\ref{silver1}, but now
multiplied by $(\varepsilon-\varepsilon_F)^2$. The dashed line represents the
$G^0W^0$ calculation for hot electrons in a FEG with $r_s=3.02$.}
\end{figure}

Fig.~\ref{silver2} shows first-principles $G^0W^0$ calculations of the
scaled lifetime $\tau(\varepsilon)\times (\varepsilon-\varepsilon_F)^2$ 
of hot electrons
in Ag, as obtained from Eq.~(\ref{eq:taudiag}) by (i) assigning the $4d$
shell to the core in the generation of the pseudopotential (squares), (ii)
including all $4d^{10}$ states only in the evaluation of ${\rm
Im}\left[\epsilon_{{\bf G},{\bf G}}({\bf q},\omega)\right]$ (triangles), (iii)
including all $4d^{10}$ states also in the evaluation of $\left|\epsilon_{{\bf
G},{\bf G}}({\bf q},\omega)\right|^{-2}$ (diamonds), and (iv) fully accounting
for the presence of $4d^{10}$ states (open circles). First
of all, we note that it matters very little whether $d$ electrons are
included in the evaluation of ${\rm Im}\left[\epsilon_{{\bf
G},{\bf G}}({\bf q},\omega)\right]$ or not. This is obviously due to the
fact that in the case of Ag and at the hot-electron energies under study, $d$
electrons are too far below the Fermi level to participate in the
creation of e-h pairs. Nevertheless, it does not matter how far they are
located below the Fermi level for them to participate in the screening of e-e
interactions. Indeed, Fig.~\ref{silver2} shows that the screening of $d$
electrons, which enters in the evaluation of $\left|\epsilon_{{\bf G},{\bf
G}}({\bf q},\omega)\right|^{-2}$, is responsible for the hot-electron lifetimes
in Ag being $\sim 2.5$ times larger than in the absence of occupied $d$ states.

\subsection{Gold}

\begin{figure}
\vspace* {1cm}
\includegraphics[width=0.95\linewidth]{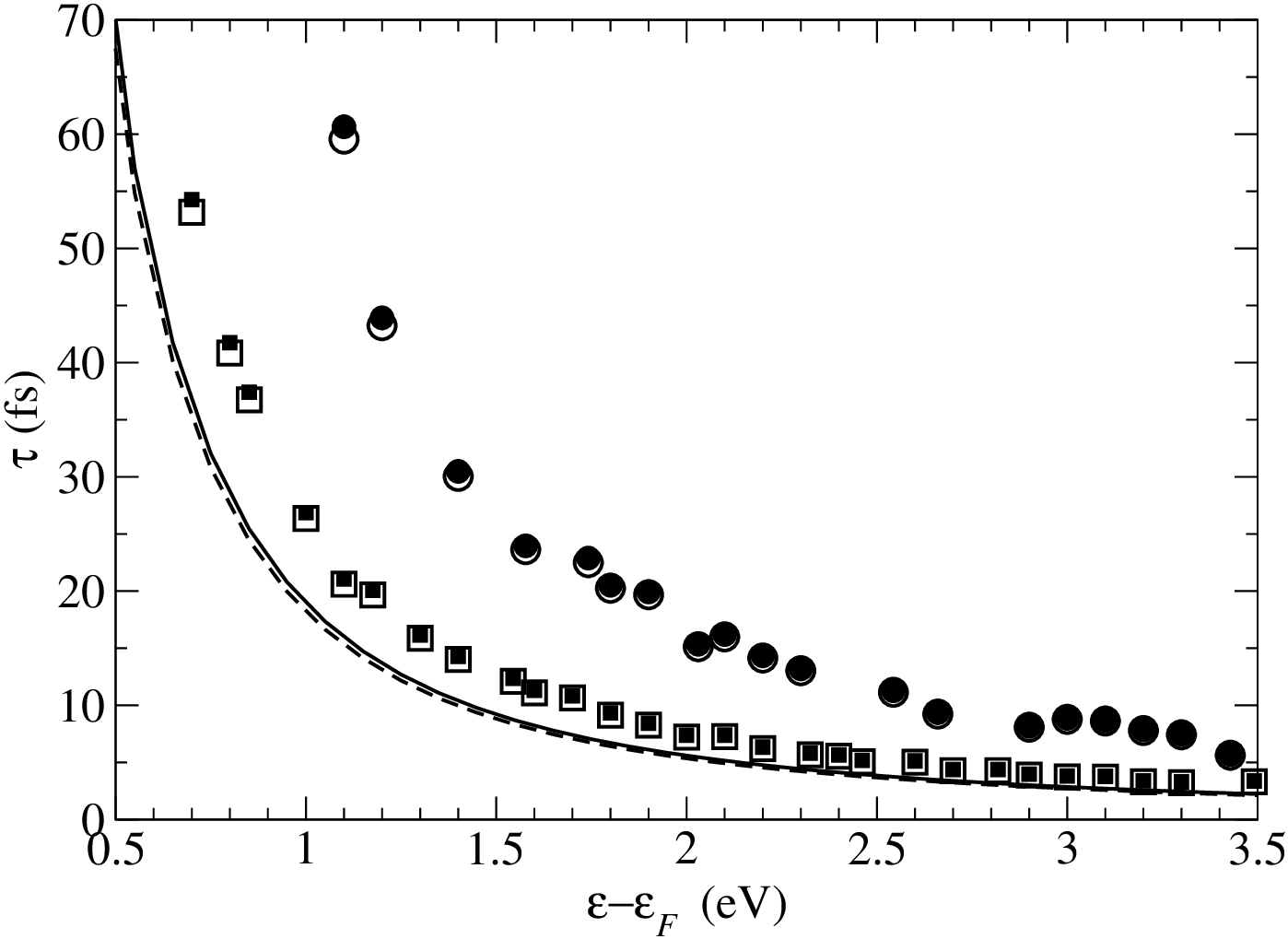}
\caption{\label{gold1}
As in Fig.~\ref{fig:cudslfe}, but for Au. Here, solid and dashed lines
represent $GW\Gamma$ and $G^0W^0$ lifetimes of hot electrons in a FEG with
$r_s=3.01$, corresponding to the average density of $6s^1$ electrons in Au.}
\end{figure}

Fig.~\ref{gold1} shows our full first-principles $GW\Gamma$ calculation
of the average lifetime $\tau(\varepsilon)$ of hot electrons in Au (solid
circles), as
obtained in the ALDA from Eqs.~(\ref{eq:tauG})-(\ref{bands3}) but with the 
inverse dielectric
matrix $\epsilon_{{\bf G},{\bf G}'}^{-1}({\bf q},\omega)$ replaced by that of
Eq.~(\ref{eq:WGG2}). For comparison, we also plot in the same figure our full
first-principles $G^0W^0$ calculation (open circles), first-principles
$GW\Gamma$ and $G^0W^0$ $s$-calculations (solid and open squares), and
$GW\Gamma$ and $G^0W^0$ FEG calculations with the electron density $n_0$ equal
to that of $6s^1$ electrons in Au (solid and dashed lines). The converged
$G^0W^0$ calculation represented in Fig.~\ref{gold1} by open circles replaces
the
lifetimes reported in Ref.~\onlinecite{igorau00}, which were too large by an
overall factor of $\sim 1.4$.

As in the case of Cu and Ag, the overall impact of exchange and correlation on
the lifetime of hot electrons in Au is very small, now $\sim 2\%$. We also
note that our pseudopotential $G^0W^0$ calculations are in excellent agreement
with the all-electron linearized augmented plane wave (LAPW) $G^0W^0$
calculations reported in Ref.~\onlinecite{au-graz}, although they are $\sim
20\%$ larger than the corresponding all-electron linear-muffin-tin-orbital
(LMTO) calculations reported in Ref.~\onlinecite{zhukovag01}. As noted in
Ref.~\onlinecite{au-graz}, this discrepancy should be attributed to the
atomic-sphere approximation used in the LMTO calculations of
Ref.~\onlinecite{zhukovag01}. The LAPW $G^0W^0$ lifetimes of
Ref.~\onlinecite{au-graz} were found to accurately reproduce the BEES spectra
for the two prototypical Au/Si and Pd/Si systems, although they were
approximately $40\%$ shorter than the TR-2PPE measurements reported in
Ref.~\onlinecite{cao98}. Our $GW\Gamma$ calculations indicate that the
inclusion of exchange and correlation does not change these conclusions.

\begin{figure}
\vspace* {1cm}
\includegraphics[width=0.95\linewidth]{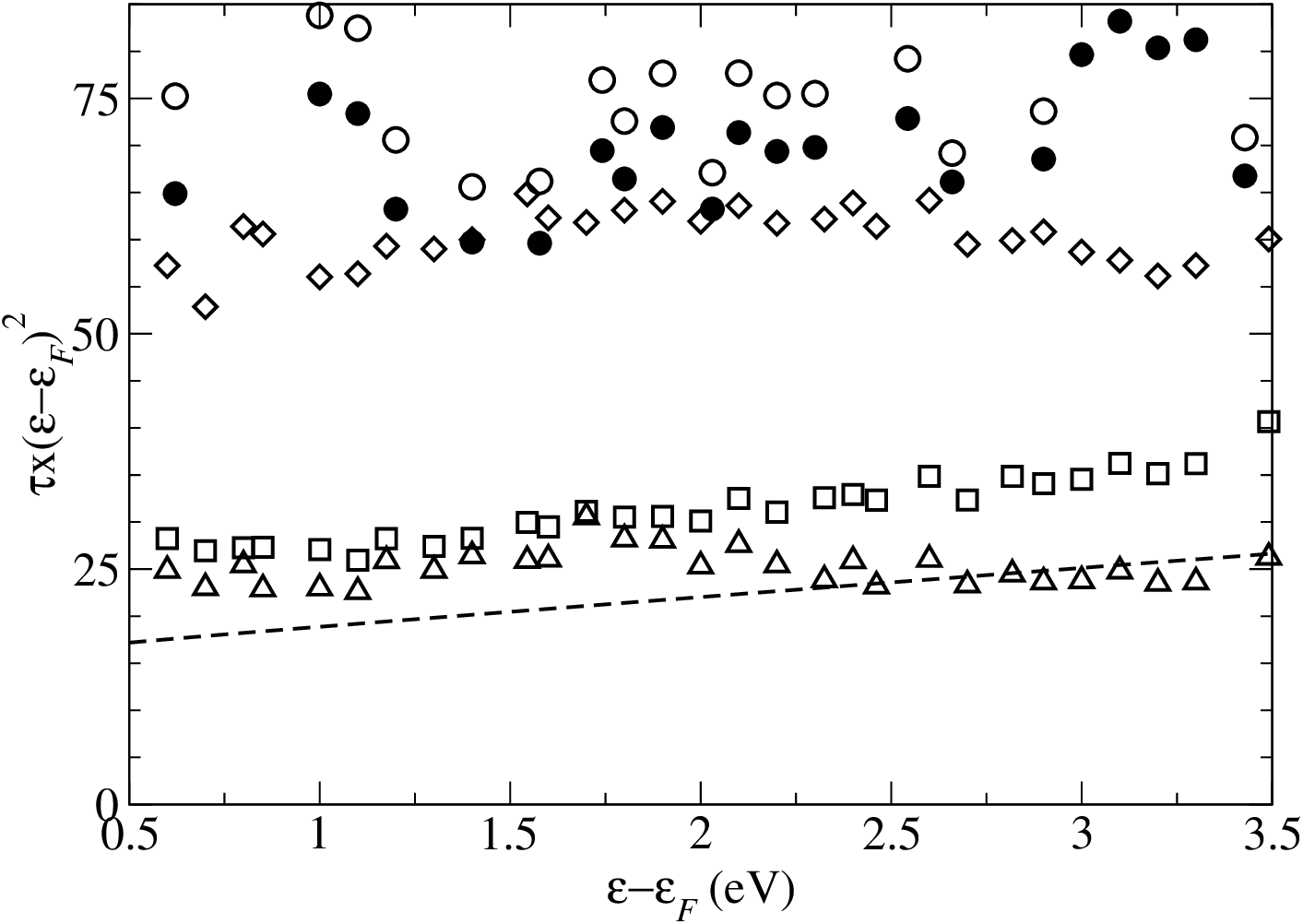}
\caption{\label{gold2}
As in Fig.~\ref{fig:cumix}, but for Au. Solid circles now represent the
full $GW\Gamma$ matricial calculation of Fig.~\ref{gold1}, but now
multiplied by $(\varepsilon-\varepsilon_F)^2$. The dashed line represents the
$G^0W^0$ calculation for hot electrons in a FEG with $r_s=3.01$.}
\end{figure}

A closer analysis of our $G^0W^0$ calculations is presented in
Fig.~\ref{gold2}, where various calculations of the scaled lifetime
$\tau(\varepsilon)\times (\varepsilon-\varepsilon_F)^2$ of hot 
electrons in Au is exhibited, as obtained
from Eq.~(\ref{eq:taudiag}) by (i) assigning the $5d$
shell to the core in the generation of the pseudopotential (squares), (ii)
including $d$ states only in the evaluation of ${\rm
Im}\left[\epsilon_{{\bf G},{\bf G}}({\bf q},\omega)\right]$ (triangles), (iii)
including $d$ states also in the evaluation of
$\left|\epsilon_{{\bf G},{\bf G}}({\bf q},\omega)\right|^{-2}$ (diamonds),
and (iv) including $d$ states everywhere (open circles). As in the
case of Cu and Ag, these calculations show the major importance of the
dynamical screening of $d$ electrons, which is responsible for the
hot-electron lifetimes in Au being $\sim 2.5$ times larger than in the absence
of
occupied $d$ states.

As noted in Ref.~\onlinecite{igorau00}, the role that occupied $d$ states play
in the screening of e-e interactions is more important in Au than in Cu,
although the static polarizable background of $d$ electrons in these materials
should be expected to be the same. However, our calculations indicate that $5d$
bands in Au are more free-electron-like than $3d$ bands in Cu, thereby
allowing Au $5d$ electrons to be more effective in the dynamical screening of
e-e interactions.

\section{\label{sum}Summary and Conclusions}

We have carried out extensive first-principles calculations of the inelastic
lifetime of low-energy electrons in the noble metals Cu, Ag, and Au, in the
framework of the $GW\Gamma$ approximation of many-body theory. This
approximation treats on the same footing xc effects between pairs of electrons
within the Fermi sea (screening electrons) and between the excited hot
electron and the Fermi sea. Our ALDA calculations indicate that the impact of
exchange and correlation on either screening or the bare interaction of the
hot electron with the Fermi sea, which is significant in the absence of $d$
states, is considerably reduced in the presence of $d$ states. Moreover, these
small xc contributions have opposite signs and it is the latter which
dominates, leading to $GW\Gamma$ lifetimes that are larger than their $G^0W^0$
counterparts by $\sim 3\%$ in Cu and $\sim 2\%$ in Ag and Au.

We have established the role that occupied $d$ states play in the relaxation of
hot electrons in the noble metals. We have found that deviations from the
hot-electron lifetimes in these materials from those of hot electrons in the
corresponding free gas of valence $sp$ electrons are mainly due to
the participation of occupied $d$ states in the screening of e-e interactions,
no matter whether occupied $d$ states can participate (as occurs in Cu and Au)
or not (as occurs in Ag) in the creation of e-h pairs. Au $5d$ electrons lie
further away from the nuclei than  Cu $3d$ electrons, thereby
occupied $d$ states in Au being more capable to screen the e-e interactions
than in Cu. The dynamical screening of $d$ electrons yields lifetimes
of hot electrons  that are larger than in the absence of $d$
states by a factor of $\sim 2$ in the case of Cu, and by a factor of $\sim 2.5$
in the case of Ag and Au.

We have found that our  $G^0W^0$ calculations of the hot-electron
lifetimes in Au are in excellent agreement with the corresponding all-electron
LAPW calculations reported in Ref.~\onlinecite{au-graz}, which gives us
confidence in the accuracy of our pseudopotential calculations. Our $G^0W^0$
lifetimes are found to be $\sim 20\%$ larger than the corresponding 
all-electron LMTO calculations reported in Ref.~\onlinecite{zhukovag01}, which
should be attributed to the atomic-sphere approximation used within the LMTO
scheme. We also note that our $G^0W^0$ lifetimes of hot electrons in the noble
metals are systematically higher than those reported in
Ref.~\onlinecite{keyling00}, especially at the lowest energies. These authors
obtained the lifetime broadening as the full width at half maximum of the
so-called spectral function, which they calculated in the $G^0W^0$
approximation. Apart from a renormalization factor which accounts for the
deviation of the hot-electron energy from its noninteracting counterpart and
which
increases the lifetime by a factor of $\sim 20\%$, the lifetimes obtained in
this way should agree with our $G^0W^0$ calculations.

Since our $GW\Gamma$ calculations are very close to their $G^0W^0$
counterparts, the inclusion of exchange and correlation does not substantially
change the comparison between $G^0W^0$ calculations and TR-2PPE and BEES
measurements existing in the literature for Cu and Au. In the case of Ag, we
have compared our new $G^0W^0$ and $GW\Gamma$ calculations with the recent
experimental TR-2PPE measurements reported in Ref.~\onlinecite{aes02}. At the
lowest hot-electron energies there is good agreement between our calculations
and the experimental data. At larger energies, our calculations are found to
lie slightly higher than the experimental curve. 

\begin{acknowledgments}
We acknowledge partial support by the University of the Basque
Country, the Basque Hezkuntza, Unibertsitate eta Ikerketa Saila, the Spanish
Ministerio de Ciencia y Tecnolog\'\i a, and the Max Planck Research Funds.
\end{acknowledgments}

\bibliography{life}

\end{document}